\documentclass[twocolumn,showpacs,preprintnumbers,amsmath,amssymb]{revtex4}
\usepackage{graphicx}
\begin{document}


\title{Tunneling spectroscopy of Na$_{x}$CoO$_2$ and Na$_{x}$CoO$_2$$\cdot$ yH$_2$O}

\author{L. Shan$^1$, H. Gao$^1$, Y.G. Shi$^1$, H.P. Yang$^1$, X.F. Lu$^1$, G.H. Cao$^2$, Z.A. Xu$^2$}
\author{H. H. Wen$^1$} \email{hhwen@aphy.iphy.ac.cn}

\affiliation{$^1$ National Laboratory for Superconductivity,
Institute of Physics, Chinese Academy of Sciences, P.~O.~Box 603,
Beijing 100080, P.~R.~China}

\affiliation{$^2$ Department of Physics, Zhejiang University, Hangzhou 310027, P.R.
China}

\date{\today}

\begin{abstract}
The single electron tunneling spectroscopy on superconductor Na$_{x}$CoO$_2$$\cdot$
yH$_2$O and its starting compound Na$_{x}$CoO$_2$ has been studied with point-contact
method. The spectra of Na$_{x}$CoO$_2$ have two types of distinct shapes at different
random locations, this is attributed to the non-uniformly distributed sodium escaped from
the inner part of the sample. While all the measured spectra of the superconducting
samples Na$_{x}$CoO$_2$$\cdot$ yH$_2$O have a good spatial reproducibility, and show a
remarkable zero bias conductance depression appearing below an onset temperature which
associates very well with the resistance upturn at around 45 K. The latter behavior
resembles in some way the pseudogap feature in high-T$_c$ cuprate superconductors.

\end{abstract}


\maketitle

After many years of intensive research on the physics of high T$_c$ cuprate
superconductors (HTSCs), there are still some crucial points needed to be clarified. A
significant one is to determine the form of any other ground state coexisting and
competing with superconductivity \cite{KangHJ2003,AlffL2003}, which is related to the
possible existence and origin of the pseudogap---a suppression of the electronic states
at the Fermi level in the normal state of HTSCs \cite{TallonJL2001,TimuskT1999}.
Nonetheless, it is now widely accepted that the physics of HTSCs is that of the doped
Mott insulator. Note that the recent synthesized superconductor Na$_{0.35}$CoO$_2$$\cdot$
yH$_2$O may also be viewed as a Mott insulator with electron doping of $35\%$ while its
T$_c$ is only about 5K \cite{TakadaK2003}. Furthermore, the superconductivity phase
diagram of Na$_{x}$CoO$_2$$\cdot$ yH$_2$O system is found to be analogy to the HTSCs
\cite{SchaakRE2003}. Therefore, the investigation on the electronic and magnetic behavior
of these materials may highlight the characteristics responsible to the high T$_c$
superconductivity of the cuprates. Besides the extensive experimental
\cite{FooML2003,SakuraiH2003,LorenzB2003,CaoGH2003,KobayashiY2003,UelandBG2003,LynnJW2003}
and theoretical \cite{BaskaranG2003, KumarB2003,WangQH2003,OgataM2003,
TanakaA2003,SinghDJ2003,SanoK2003,LiT2003,SaD2003} studies of this new superconducting
system, its starting compound Na$_{x}$CoO$_2$ has also attracted much attentions
\cite{TerasakiI1997,TanakaT1994,RayR1999,OnoY2001,HasanMZ2003} and has been demonstrated
to have some unconventional transport behavior and magnetic transition when $x=0.75$
\cite{MotohashiT2003}. However, to our knowledge no spectroscopic studies of both
Na$_{x}$CoO$_2$$\cdot$ yH$_2$O and Na$_{x}$CoO$_2$ have been reported so far in
literature.

In this paper, we present the results of point-contact spectroscopy measurements on
Na$_{x}$CoO$_2$ and Na$_{x}$CoO$_2$$\cdot$ yH$_2$O compounds. The spectra of these two
kinds of compounds have some essential differences not only on the spectral shape but
also on their temperature dependence and spatial uniformity. Some possible mechanisms are
suggested to explain all the observed novel spectral characteristics.

\begin{table}
\caption{\label{tab:table1} Sample information of the superconducting
Na$_{x}$CoO$_2$$\cdot$ yH$_2$O ($a$ and $c$ are lattice parameters).}
\begin{ruledtabular}
\begin{tabular}{cccccc}
sample & $x$  & $a$($\AA$)  & $c$($\AA$) & $y$ & T$_c$(K)\\
\hline
SC026      & 0.26    &2.821     &19.807    &$\approx1.3$ &3.5\\
SC031      & 0.31    &2.820     &19.650    &$\approx1.3$ &4.1\\

\end{tabular}
\end{ruledtabular}
\end{table}

\begin{figure}[]
\includegraphics[scale=1.2]{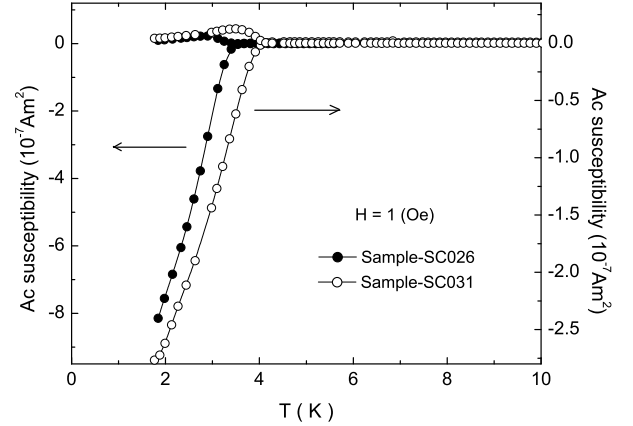}
\caption{\label{fig:fig1-Tc} Superconducting transitions measured by Ac susceptibility. }
\end{figure}

The information of the Na$_{x}$CoO$_2$$\cdot$ yH$_2$O samples used in this work are
presented in Table~\ref{tab:table1}. Fig.~\ref{fig:fig1-Tc} shows the superconducting
transition characterized by AC susceptibility measurements. The starting compounds of
these two samples is Na$_{0.70\pm 0.05}$CoO$_2$ (NSC070). Details of the sample
preparation and characterizations have been given in Ref.\cite{ShiYG2003} for
Na$_{0.26}$CoO$_2$$\cdot$ 1.3H$_2$O (SC026) and in Ref.\cite{CaoGH2003} for
Na$_{0.31}$CoO$_2$$\cdot$ 1.3H$_2$O (SC031). Powder X-ray Diffraction measurements
indicate that both samples are hexagonal single phase. The spectroscopy measurements on
these samples (SC026,SC031 and NSC070) were proceeded with N/S (N=PtIr alloy; S denote
samples) point contacts in the sample chamber of an Oxford cryogenic system MaglabExa12.
As described in our previous works \cite{ShanL2003}, the PtIr tips were prepared by
electrochemical etching in CaCl$_2$ solution using Pt$_{0.9}$Ir$_{0.1}$ wire with a
diameter of 0.25 mm. The approaching of the tip toward the sample was controlled by a
differential screw. Typical four-terminal and lock-in technique were used to measure the
differential resistance $dV/dI$ vs $V$ of the point contacts. Then the $dV/dI-V$ curves
were converted into the dynamical conductance $dI/dV-V$ (or $\sigma-V$) curves. In our
tunneling spectroscopy measurements, we have processed bidirectional bias-voltage
scanning in various rates, all the spectra presented in this paper are reproducible and
have no hysteresis, so that the heating effect can be ruled out in these data.

\begin{figure}[]
\includegraphics[scale=1.2]{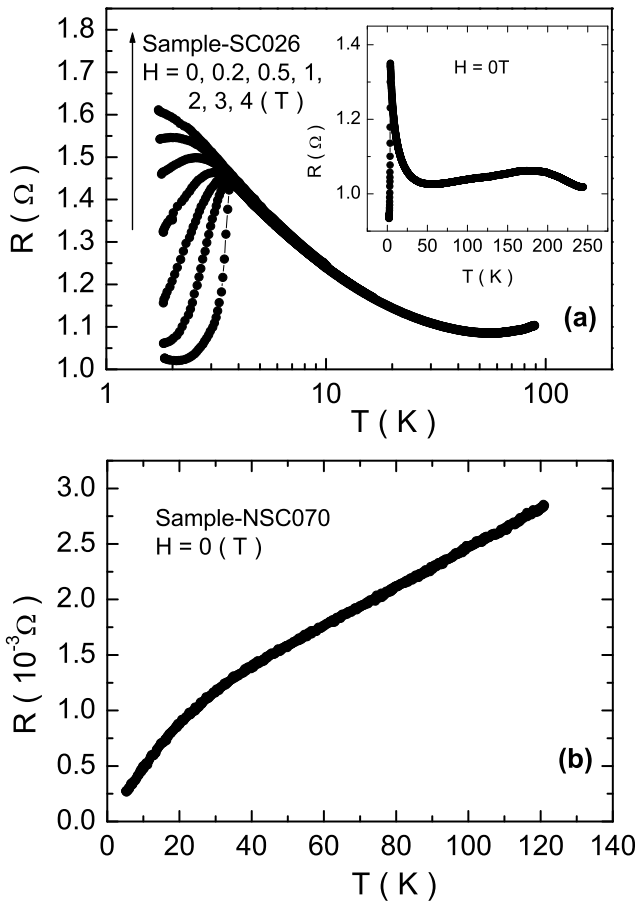}
\caption{\label{fig:fig2-RT} (a) Temperature dependence of resistance of sample-SC026 at
various magnetic fields from 0T to 4T. Inset is the large scale data measured at 0T. It is clear that a low temperature upturn of resistance appears at around 50 K. (b) Temperature dependence of resistance of sample-NSC070 measured at 0T.}
\end{figure}

Fig.~\ref{fig:fig2-RT}(a) shows the temperature dependence of the resistance of
sample-SC026 at various magnetic fields. The superconducting transition becomes broad
with increasing field and an increase of normal-state resistance with decreasing
temperature occurs at about 50K. However, the onset T$_c$ does not depend significantly
on the magnetic field. These characteristics are the same as that of sample-SC031 and the
results presented in Ref.\cite{TakadaK2003}. Similar behavior is always observed in
high-T$_c$ copper oxides. As a comparison, we present in Fig.~\ref{fig:fig2-RT}(b) the
R-T relation of sample-NSC070. It is noted that the resistance decrease monotonically
with decreasing temperature and a weak inflexion occurs at around 30K. Such remarkable
resistive difference between the superconductors and their starting compounds with similar
structure should be a reflection of their different electronic states.

\begin{figure}[]
\includegraphics[scale=1.2]{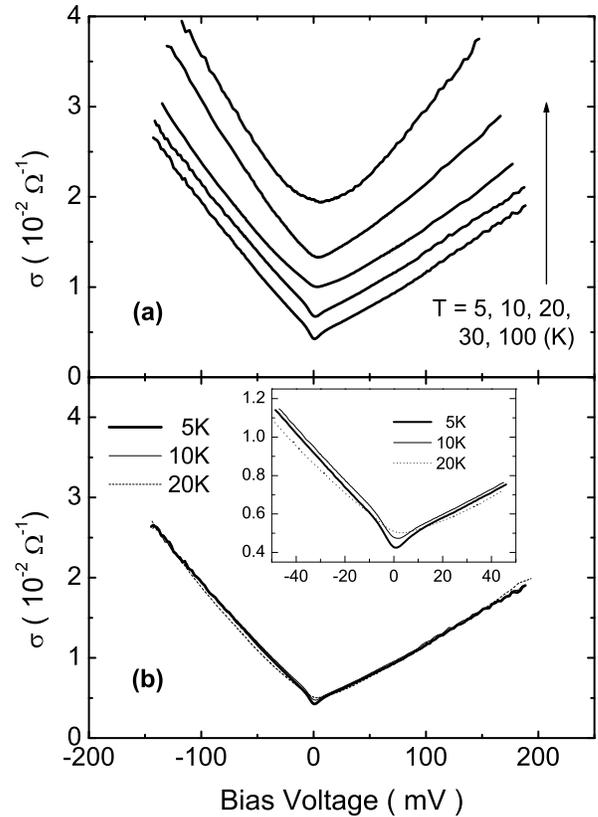}
\caption{\label{fig:fig3-TSnsc} Typical tunneling spectra of PtIr/NSC070 point contact
measured at a fixed point on the sample surface for various temperatures. (a) All curves
except the bottom one have been shifted upwards for clarity. (b) Raw data of three
consecutive temperatures. Inset shows the enlarged detail around zero bias voltage.}
\end{figure}

Fig.~\ref{fig:fig3-TSnsc} shows the typical tunneling spectra measured on the PtIr/NSC070
point contacts. Examining the gross features of a tunneling conductance curve shown in
Fig.~\ref{fig:fig3-TSnsc}, one notes the background with a slight positive curvature,
which is ordinary for the metal-insulator-metal junctions. A closer scrutiny in the
region near zero bias reveals a small conductance dip as shown in the inset of
Fig.~\ref{fig:fig3-TSnsc}(b). Such small dips exist between a narrow bias range of $\pm
10$mV at low temperatures while disappears for higher temperatures. These features are
similar to that of nonequilibrium electron tunneling in metal-insulator-metal junctions
\cite{AlderJG1975,WolfEL1985}.

\begin{figure}[]
\includegraphics[scale=1.2]{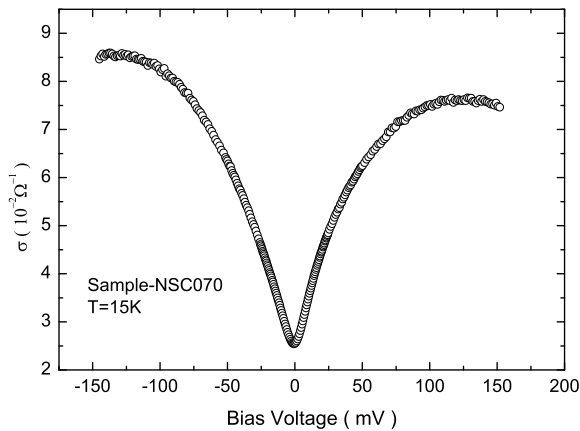}
\caption{\label{fig:fig4-TSspecial} The other type of spectra of PtIr/NSC070 point
contact observed occasionally, which is associated with Giaever-Zeller two step tunneling
as discussed in the text. }
\end{figure}

As shown in Fig.~\ref{fig:fig4-TSspecial}, the other type of spectrum of sample-NSC070
has also been observed occasionally at some random positions of the sample surface.
However, the little probability of observing such spectrum and its bad reproducibility
indicate that it comes from some local properties. The zero bias conductance dip shown in
Fig.~\ref{fig:fig4-TSspecial} is much more significant than that shown in
Fig.~\ref{fig:fig3-TSnsc}, that is, a giant resistance peak occurs. According to the
model proposed by Zeller and Giaever \cite{ZellerHR1969}, such giant resistance peak
around zero bias can be achieved if tunneling occurs through intermediate states on metal
particles embedded in the oxide barrier. This is very close to the surface condition of
sample-NSC070, since the sodium atoms between the CoO$_2$ layers in Na$_x$CoO$_2$ are
easy to escape from the bulk. Because the small amount of separated sodium on the surface
are easily oxidated, it is possible in some positions of sample surface to form the oxide
(or carbonate) barrier with embedded sodium particles, and hence the Giaever-Zeller two
step tunneling model is favored. A direct consequence of such spatial un-uniformity is
that, when the tip locates closely to the phase boundaries of the normal regions and the
Giaever-Zeller regions, even a slight tip relaxation caused by temperature variation can
lead to a remarkable change of the measured spectra. This is indeed observed in few cases
in our measurements.

\begin{figure}[]
\includegraphics[scale=1.2]{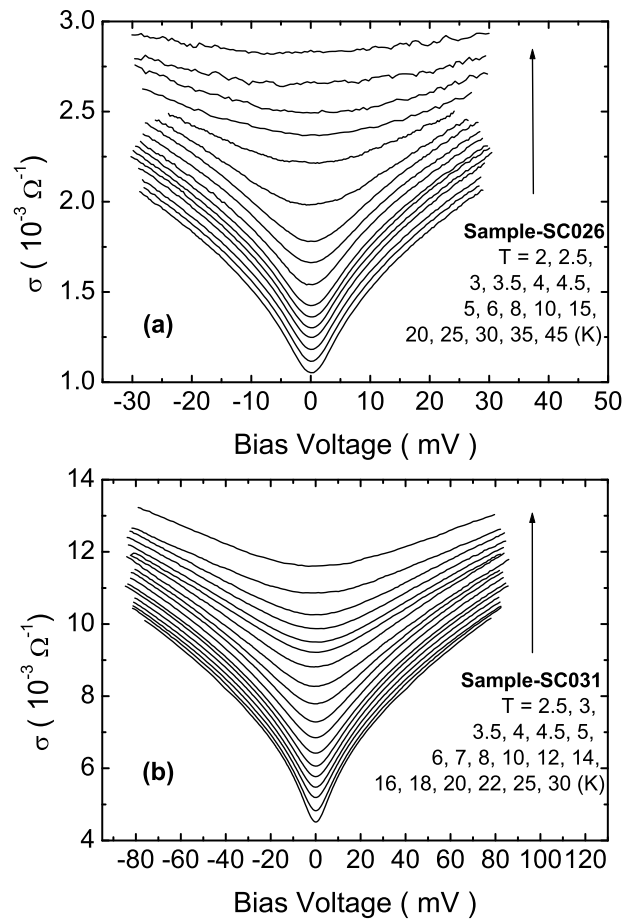}
\caption{\label{fig:fig5-TSsc} Temperature dependence of the tunneling spectra of
PtIr/SC026 (a) and PtIr/SC031 point contacts. All curves except the bottom one has been
shifted upwards for clarity. Noting the similar spectral shape and temperature dependence
of these two samples. }
\end{figure}

The measured tunneling spectra of samples SC026 and SC031 are presented in
Fig.~\ref{fig:fig5-TSsc}(a) and (b) respectively. No characteristic of superconductivity
such as coherence peak can be detected in the measured tunneling spectra. This may be
ascribed to the very low T$_c$, broad superconducting transition, and extremely low
density of charge carriers. However, by comparing these results with that of NSC070, we
can find following remarkable differences between them besides their distinct spectral
details. Firstly, the spectra of superconducting samples have a good symmetry for the
opposite sides of zero bias while the non-superconducting sample does not; Secondly, the
spectral shape is identical for various positions on the surfaces of the superconducting
samples, unlike their starting compounds. The good reproducibility of the temperature
dependent spectra for the samples SC026 and SC031 in several thermal cycles indicates the
phase stability and spatial homogeneity of the electronic states. The spatially
consistent conductance depression may rule out the spatial-related mechanisms such as
above discussed Giaever-Zeller model and the magnetic impurity effect. A roughly
estimation of the onset temperature of this characteristic is somewhat above 45K for
sample-SC026, which is close to the onset temperature of the resistance upturn as shown
in Fig.~\ref{fig:fig2-RT}(a). Another explanation for this conductance depression is the
pseudogap effect as observed in cuprate high T$_c$ superconductors. In fact, the shape
and temperature dependence of this conductance depression are similar to that observed in
the electronic doped cuprate superconductor Pr$_{2-x}$Ce$_x$CuO$_{4-y}$
\cite{AlffL2003,BiswasA2001}. In this sense, the occurence of the conductance deppression
and the resistance enhancement with decreasing temperature correspond to the opening of
the pseudogap around the Fermi level.

\begin{figure}[]
\includegraphics[scale=1.2]{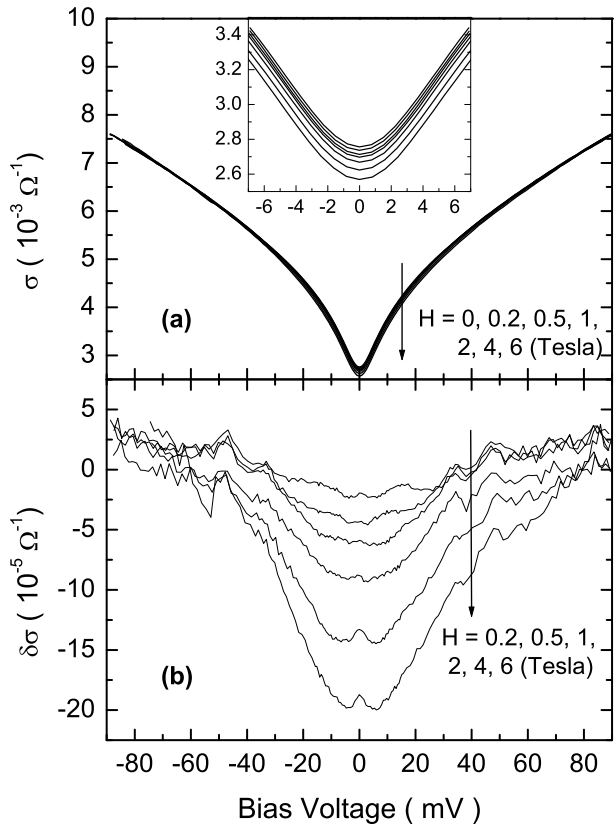}
\caption{\label{fig:fig6-field} (a) Magnetic field dependence of the tunneling spectra of
PtIr/SC031 point contact. Inset shows the enlarged details around zero bias voltage. (b)
Redraw the data as show in (a) after a subtraction of the data of 0T. }
\end{figure}
In order to get further insight into this anomalous conductance depression, we have
studied the magnetic field influence on the tunneling spectroscopy.
Fig.~\ref{fig:fig6-field}(a) shows the field dependence of the tunneling spectra and the
details around zero bias are enlarged in the inset. It is noted that the spectra measured
at various fields overlap each other at high bias while gradually separate from each
other with decreasing bias voltage. In order to give a more clear elucidation, we present
in Fig.~\ref{fig:fig6-field}(b) the field dependent spectra after a subtraction of the
zero field one. It is obvious that the applied magnetic field induces an additional zero
bias conductance depression with a magnitude in proportional to the magnetic field.

In summary, we have performed tunneling spectroscopy measurements on the
PtIr/Na$_{x}$CoO$_2$ and PtIr/Na$_{x}$CoO$_2$$\cdot$ yH$_2$O point contacts. The observed
spatially sensitive spectra of Na$_{x}$CoO$_2$ are associated with the non-uniformly
distributed escaped sodium and its oxide. While the spatially consistent conductance
depression in the spectra of Na$_{x}$CoO$_2$$\cdot$ yH$_2$O seems to be related to the
resistance upturn at low temperature, which resembles in some way the pseudogap effect
observed in the underdoped cuprate superconductors. However, its anomalous magnetic field
dependence can not be directly comparable to the cuprate superconductors. Additional
experiments are still needed to clarify this issue.

\begin{acknowledgments}
This work is supported by the National Science Foundation of China (NSFC 19825111,
10074078), the Ministry of Science and Technology of China ( project: NKBRSF-G1999064602
), and Chinese Academy of Sciences with the Knowledge Innovation Project.
\end{acknowledgments}


\end{document}